\documentclass[useAMS]{mn2e}
\usepackage{graphicx} 
\usepackage{epstopdf}

\title[3C 345: the historical light curve (1967-1990) from the digitized plates 
of the Asiago Observatory]{3C 345: the historical light curve (1967-1990) from the digitized plates 
of the Asiago Observatory}
\author[A. Omizzolo, C. Barbieri, C. Rossi]{A. Omizzolo$^{1}$\thanks{E-mail:
omizzolo@pd.astro.it}, C. Barbieri $^{2}$ and C. Rossi$^{3}$ \\
$^{1}$ Specola Vaticana, Citt\`a del Vaticano\\
$^{2}$Dipartimento di Astronomia, Universit\`a di Padova, 
Vicolo dell'Osservatorio 5, 35100 Padova, Italy\\
$^{3}$Dipartimento di Fisica, Universit\`a La Sapienza, 
Piazzale Aldo Moro 3, 00185 Roma, Italy} 

\begin{document}

\date{Accepted . Received ; in original form }

\pagerange{\pageref{firstpage}--\pageref{lastpage}} \pubyear{2004}

\maketitle

\label{firstpage}

\begin{abstract}
In the frame of a large project to digitize the plate archives of the 
Italian and Vatican Astronomical Observatories, we have already performed the 
digitization of all available plates of the field of the quasar 3C345. The plates, 
approximately 100, were taken with the three telescopes of the Asiago Observatory 
(122 cm, 182 cm, 67/90 cm Schmidt Telescope) in the period from 1967 to 1990. We present here essentially new data, mostly in the B band, about the variability of 3C 345 and also of other four objects (3 quasars and the active galaxy NGC 6212) in the same field, in that period. Beyond the well known 3C 345 itself, also the other three quasars show variability, with a range of 2.0 mag for 
Q1 and Q2, 1 mag for Q3.  The low level variability detected for the nucleus of NGC 6212 is more suspicious, and should be confirmed by linear detector data. 
\end{abstract}


\section{Introduction}

The interest in the long time scale variability of quasars has been recently revived thanks to new optical, X and radio data. The interest of long-term monitoring programs is discussed for istance by Hawkins (2002). Well followed in particular is 3C 345, one of the most luminous and violently variable ones, for which many recent studies have highlighted the complex behavior (Caproni et al., 2004, Zhang et al., 
2000). Many plates of 3C 345 were taken at the Asiago Observatory in a larger 
project aimed to study the optical variability of a selected sample of quasars 
(see for instance Barbieri et al., 1988 and references therein). Only part of the 
Asiago data have been published (Barbieri et al. 1977 a,b), but many exposures 
were taken after that publication. Here we present the magnitudes and the light 
curve covering the entire period of the Asiago observations; while the paper of 
1977 gave magnitudes estimated by traditional means on the original photographic 
emulsion, here we publish the results of the photometric analysis of the digitized 
plates. This work is one of the first scientific results of the National project to 
preserve the photographic archives of several Italian Observatories and of the 
Vatican Observatory (see Barbieri et al., 2003).

\section{The field of  3C 345}
The plates of the 3C 354 field were taken mainly in the B band (103aO+GG13, or 
IIaO+GG13), with few plates also in V (103aD+GG11, 103aD+GG14) and U (103aO+UG2) 
using three telescopes, namely the 122 cm at its f/5 Newtonian focus, the S67/92 
cm Schmidt telescope and the 182 cm at the f/9 Cassegrain focus.
For the present paper, we have limited the analysis of the images to the field 
common to all three telescopes (namely, the field of the 182 cm telescope), where 
not only 3C345 but also other three QSOs and one active galaxy (NGC 6212) are 
reported by the Veron (Veron \& Veron 2003) Catalog. Positions and redshifts of these objects are 
presented in Table 1.
Fig. 1 shows the field of 3C 345 with the photometric sequences used in this work, the positions of the other four QSOs and of nine USNO stars (S1-S9) randomly chosen in the field to test the goodness of the photometric method. Two more quasars (Q5, Q6) are indeed present in the field but they are too faint to be detectable in most plates. We have only two good points, and therefore we did not consider them.

%======================= Figure 1 =========================
\begin{figure*}
\centering 
\includegraphics[width=17cm]{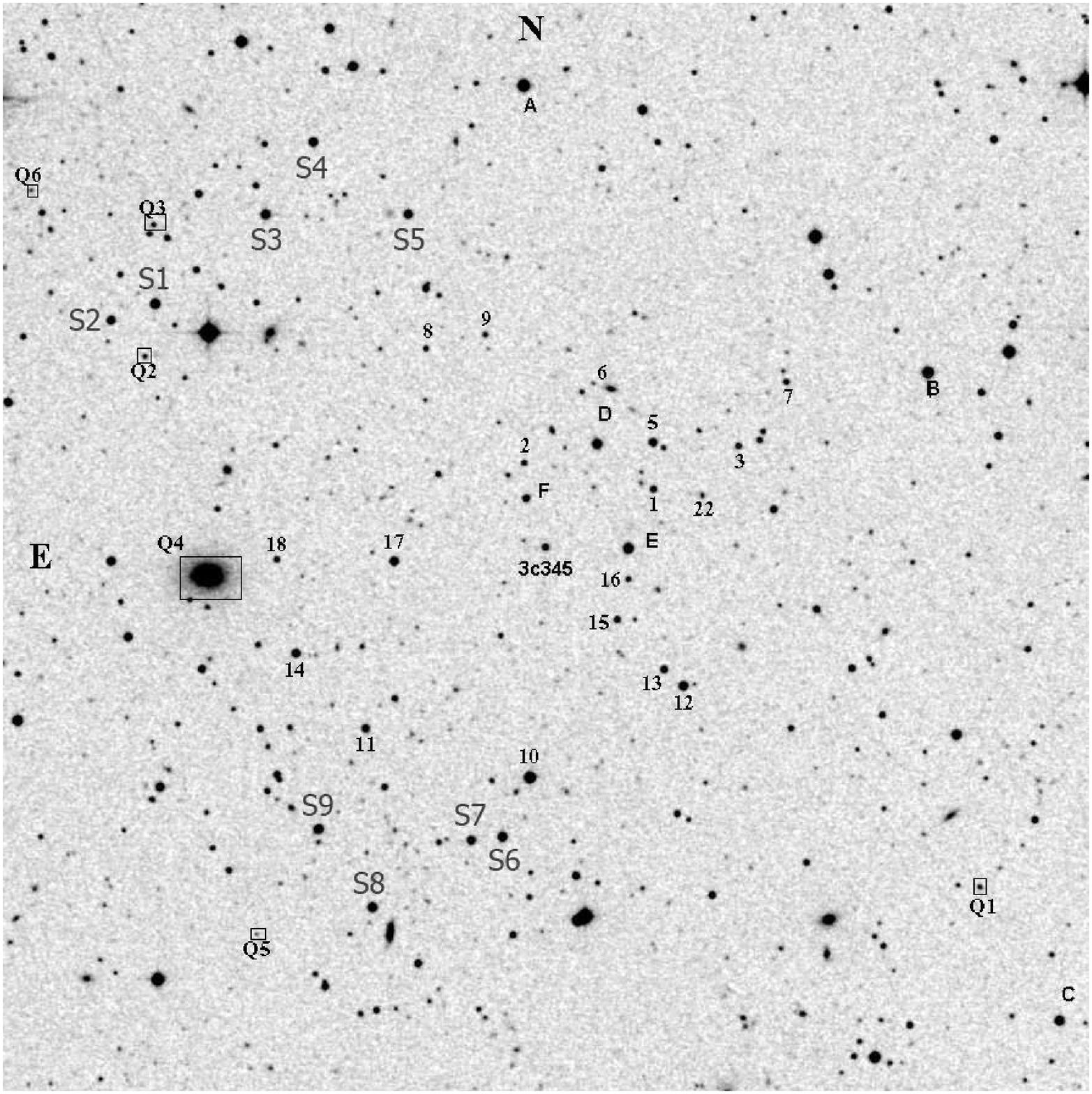} 
\caption{The field of 3C 345 and other 4 extragalactic objects, with the used 
photometric sequence and the 9 test stars (S1-S9) randomly chosen in the field.}
\end{figure*}
%============================================================

%========================== Table 1 =========================
\begin{table}
\caption[]{Positions and redshifts of the 3 QSOs and of the active galaxy NGC 6212 in the field of 3C 345}  
\begin{flushleft}
\begin{tabular}{cccccrcc}
\hline
\noalign{\smallskip} 
\noalign{\medskip} 
name  &$\alpha(2000)$   &$\delta(2000)$ &$z$ 
\\
\noalign{\medskip} 
\hline 
\noalign{\medskip} 
1640.8+398(Q1)  & 16 42 28.5 & +39 43 43 & 1.86\\
3C 345          & 16 42 58.8 & +39 48 37 & 0.59\\
NGC 6212 (Q4)   & 16 43 23.1 & +39 48 24 & 0.03\\ 
E1641.7+3988(Q3)& 16 43 26.2 & +39 53 14 & 0.704\\
1641.8+399(Q2)  & 16 43 27.1 & +39 51 25 & 1.083\\

\noalign{\medskip} 
\hline 
\noalign {\medskip} 
\end{tabular}
\end{flushleft} 
\end{table} 
%===============================================================

\section{The digital acquisition and the photometric reduction} 

As detailed in Barbieri et al. (2003) the plates have been digitized with a good 
quality commercial scanner at 1600 dpi in transparency mode (positive = low 
transparency for the stars, high  transparency for the sky) with 14 bit 
resolution. An "ad hoc" program allows to override the automatic scanner software 
and to write directly the resulting image in FITS format. 
The main part of the work to obtain the magnitudes of the QSOs consisted in 
setting up a standard procedure for the calibration of the scanned images, in 
order to use  standard software packages (i.e. IRAF, in particular the tasks 
DAOPHOT and those related to it).
For the reduction  from the transparencies (scanner data numbers) to intensity 
values, we applied the methods used for the photographic emulsions following the 
algorithms for the linearization of characteristic curves by de Vaucouleurs 
(1968).
For most of the plates  sensitometric spots are not available; therefore we 
approximately converted the transparencies into  the Baker's photographic density 
BD=Log((V-T)/(T-B)) and then into fluxes assuming for all the plates a standard 
density-intensity relation; here V is the average value of the unexposed plate, T is the scanner Data Number and B is the "black", a value just below the darkest pixels of the overexposed stars. 
Of the several available photometric sequences, we considered those by Angione 
(1971, in the following A) and by Gonzales-Perez and Kidger (2001, in the following GPK) with revised values kindly communicated by Kidger (private communication, 2004). Table 2 reports the synopsis of the two sequences. 
Three stars are in common to the two sequences, for which we preferred the GPK values in order to have a sequence as homogeneous as possible. We finally performed the photometric analysis of the data by using the aperture package DAOPHOT.
For each plate we derived the calibration curve, an example of which is shown in figure 2.
In this figure the errors on the instrumental magnitudes are also indicated. 
The agreement between the IRAF magnitudes and the photometric sequence is as detailed in Table 3. The three telescopes provide magnitudes having essentially the same error.
We report in fig. 3 a comparison between the magnitudes for 3C 345 as derived from 
Barbieri et al. (1977) and the magnitudes we obtained from the digitized images of the 
same plates.

%========================== Table 2 =========================
\begin{table*}
\centering
\begin{minipage}{130mm}
\caption{Synopsis of the photometric sequences by Angione and Gonzales-Perez}  
\begin{flushleft}
\begin{tabular}{cccccrccc}
\hline
\noalign{\smallskip} 
\noalign{\medskip} 
Star  &   B(GPK)   &V(GPK) &U(GPK) &B(A) &V(A) 
&U(A) & Variability & Notes\\
\noalign{\medskip} 
\hline 
\noalign{\medskip} 
1& 19.34 & 17.65 & & & &&&\\
2& 18.46 & 17.94 & 18.26 & & &&&\\
3& 19.50 &17.89  &&& \\
4,D& 16.05 & 15.29 & 16.43 & 16.08 &15.18 &16.73&&\\
5& 17.96 & 16.71 &  & & &&&\\
6& 18.95 & 18.34 &  & & &&Variable&\\
7& 18.02 & 17.65 &  & & &&&\\
8& 19.10 & 18.35 &  & & &&&\\
9& 19.54 & 18.43 &  & & &&Variable& Long period\\
10& 13.96 & 13.77 & 13.89 & & &&Variable& Low amplitude\\
11& 17.01 & 16.51 &  & & &&&\\
12& 16.05 & 15.43 & 16.14 & & &&Varable?&\\
13& 17.22 & 16.72 & 16.97 & & &&&\\
14& 17.10 & 16.08 &  & & &&&\\
15& 17.68 & 17.14 & 17.52 & & &&&\\
16& 19.49 & 18.55 &  & & &&&\\
17& 16.38 & 15.71 &  & & &&&\\
18& 19.15 & 17.99 &  & & &&&\\
19,E& 16.45 & 15.24 & 16.93 & 16.47 & 15.22 & 17.58&&\\
21,F& 17.09 & 16.52 & 16.32 & 17.17 & 16.52 &&&\\
A&  &  &  & 14.44 & 13.76 & 14.34&&\\
B&  &  &  & 14.85 & 14.14 & 14.71&&\\
C&  &  &  & 15.87 & 15.29 & 15.83&&\\ 
\noalign{\medskip} 
\hline 
\noalign {\medskip} 
\end{tabular}
\end{flushleft} 
\end{minipage}
\end{table*} 
%===============================================================

%========================== Table 3 =========================
\begin{table}
\caption[]{Standard deviation of the B magnitudes}  
\begin{flushleft}
\begin{tabular}{cccccrcc}
\hline
\noalign{\smallskip} 
\noalign{\medskip} 
Telescope  &   $\sigma$ B(mag)
\\
\noalign{\medskip} 
\hline 
\noalign{\medskip} 
S67/92& 0.14 $\pm$ 0.03\\
122& 0.17 $\pm$ 0.05\\
182& 0.15 $\pm$ 0.05\\
\noalign{\medskip} 
\hline 
\noalign {\medskip} 
\end{tabular}
\end{flushleft} 
\end{table} 
%===============================================================

%======================= Figure 2 ============================ 
\begin{figure}
\centering
\includegraphics[width=7cm]{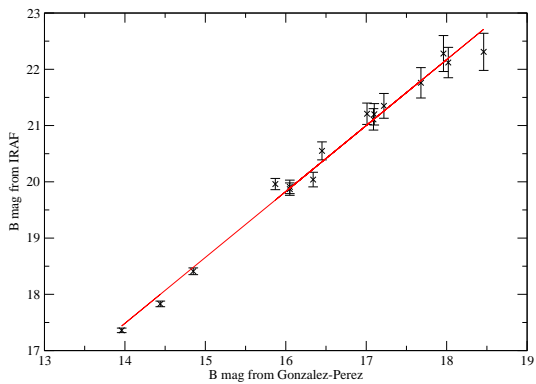} 
\caption{The fit is represented by the equation:  y = 1.0757 + 1.172 * x with an rms of $\pm 0.19$ 
mag.}
\end{figure}
%=============================================================

%======================= Figure 3 ============================ 
\begin{figure}
\centering
\includegraphics[width=7cm]{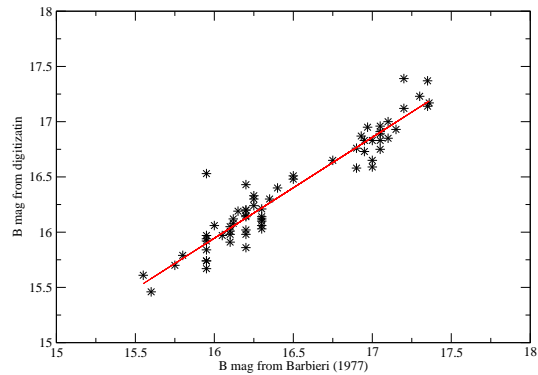} 
\caption{A comparison between the B magnitudes of 3C 345 from Barbieri 
(1977) and those obtained by us. The fit equation is y = 1.3325 + 0.91329 * x with an rms of $\pm 0.15$ mag.}
\end{figure}
%=============================================================

\section{The USNO stars}
To check the photometric reliability of the present method we performed the photometry of nine USNO stars (S1-S9), chosen in the mag interval from B=15.0 to B=17.5 on nine 120cm telescope plates. 
Their measured B magnitudes are indicated in Fig. 4 together with those of 3C 345. The plot shows that the variations of the stars never exceed $\pm$ 0.1 mag, while for 3C 345 the variation attains the value of 0.48 mag. It is also noteworthy the fact that the small variations from one plate to another for the different stars are not correlated, confirmimg that no systematic effect is introduced by the scanning procedure. The USNO magnitudes of the nine stars, their mean observed values and the corresponding mean standard deviations are reported in Table 4. For comparison, the standard deviation of the B mag of 3C 345 on this sample of plates is 0.18, several times larger than that of the nine stars; in other words, an automatic detection procedure would have singled out 3C 345 as a variable even on this small sample.

 %======================= Figure 4 ============================ 
 \begin{figure}
\centering
\includegraphics[width=7cm,angle=-90]{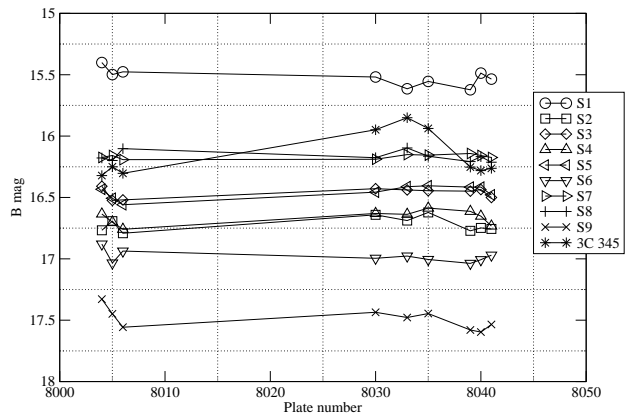} 
\caption{The Daophot photometry of 9 USNO stars and of 3C 345 on 9 120cm plates.}
\end{figure}
%=============================================================

%======================= Table 4 ============================ 
 
\begin{table*}
\centering
\begin{minipage}{100mm}
\caption[]{USNO mag, measured B mag and Standard deviation of 9 USNO stars}  
\begin{flushleft}
\begin{tabular}{ccccccccc}
\hline
\noalign{\smallskip} 
\noalign{\medskip} 
S1 & S2 & S3 & S4  & S5 & S6 & S7 & S8 & S9\\
\noalign{\medskip} 
\hline 
\noalign{\medskip} 
15.6 & 16.4 & 16.3 & 16.6 & 16.4 & 16.8 & 16.3 & 16.5 & 17.3 \\
15.53 & 16.72 & 16.46 & 16.66 & 16.45 & 16.98 & 16.17 & 16.16 & 17.48 \\
0.07 & 0.06 & 0.04 & 0.06 & 0.05 & 0.05 & 0.01 & 0.04 & 0.08\\
\noalign{\medskip} 
\hline 
\noalign {\medskip} 
\end{tabular}
\end{flushleft} 
\end{minipage}
\end{table*} 
%=============================================================

\section{The light curves}
The logbook of the available plates and the corresponding magnitudes are reported 
in Table 6 and 7. For point like objects, the measured magnitudes are independent of the used telescope; the case of the galaxy NGC 6212, that will be discussed below, shows that for extended objects it is not so (see fig. 9). The light curves in the B magnitude are shown in  figs. 5, 6, 7, 8 and 9.
As already said, some of the observations are in the U and V photometric bands.  Plates of photometric quality in all three colours in the same night are available in a few cases where the U band is always represented by only one plate. The information on the colour indexes of the objects are detailed in Table 5. Although the errors in the color indexes can be as high as $\pm$ 0.2 mag, it can be concluded that Q1, Q2 and Q3 are indeed extremely blue, while the colours of the nucleus of NGC 6212 are decidedly redder. In what follows we will discuss the risults individually for the five objects.

%======================= Table 5 ============================ 
\begin{table*}
\centering
\begin{minipage}{100mm}
\caption[]{Colour Indexes of the QSOs}  
\begin{flushleft}
\begin{tabular}{ccccccc}
\hline
\noalign{\smallskip} 
\noalign{\medskip} 
MJD&Colour Index&Q1&Q2&Q3&NGC6212&3C 345\\
\noalign{\medskip} 
\hline 
\noalign{\medskip} 
39597.06&U-B&&-0.18&-0.21&-0.83&-0.98\\
39616.06 &U-B&-1.10&-0.97&-0.50&+0.12&-0.79\\ 
39618.05&U-B&-0.94&-1.25&-0.85&+0.34&-0.63\\
39618.05&B-V&&&&+0.64&+0.17\\
39622.98&B-V&&&-0.37&+1.20&+0.18\\
39712.95&U-B&-0.46&-0.52&-0.53&+0.16&-0.57\\ 
39915.18&B-V&&&-0.14&+1.37&-0.12\\
43312.97&B-V&&&&&-0.05\\
\noalign{\medskip} 
\hline 
\noalign {\medskip} 
\end{tabular}
\end{flushleft} 
\end{minipage}
\end{table*} 
%=============================================================

\subsection{3C 345}

%======================= Figure 5 =========================
\begin{figure}
\centering 
\includegraphics[width=7cm]{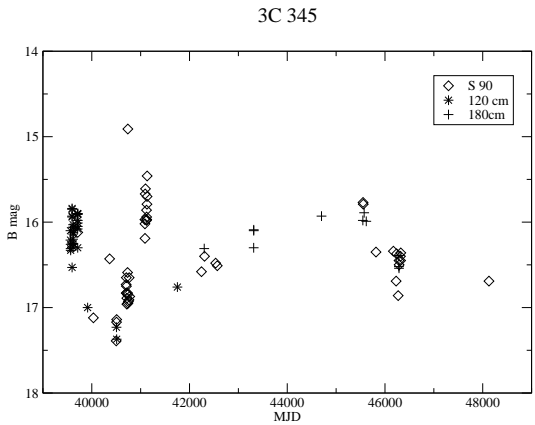} 
\caption{The light curve of 3C 345. The 3 symbols refer to the 3 used telescopes.}
\end{figure}
%============================================================
The data for 3C 345 are coherent with the behavior reported by Zhang et al.  (1998, 2000).This new Asiago light curve (Fig. 5) confirms the very complicated behaviour of the luminosity 
variations characterized by a continuous activity at the level of $\pm 0.3$ mag with superimposed sudden outbursts of approximately 2 mag.

\subsection{1640.8+398 (Q1)}
We have not found in the literature any detailed study of this faint object that shows a variability range of approximately 2 magnitudes. We report three maxima occurring at MJD 39618 (B = 18.64), MJD 41122 (B = 18.82) and MJD 42534 (B = 18.88). The first and the second happen during a period of activity 
lasting about 2 months, during which the object seems to oscillate around the 
maximum with an amplitude of 0.5 mag (see fig. 6). After 1977 the QSO was too 
faint and not detected any longer.

%======================= Figure 6 ============================ 
\begin{figure}
\centering
\includegraphics[width=7cm]{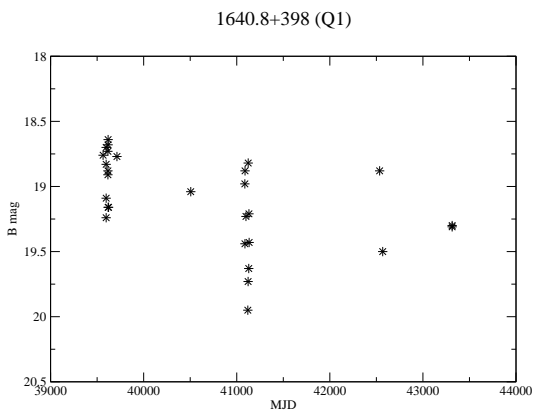} 
\caption{The light curve for 1640.8+398 (Q1)}
\end{figure}
%=============================================================
\subsection{1641.8+399 (Q2)}
As for Q1, we did not find data about this object in the literature. Our data 
cover the period from 1967 to 1985 with a total of 42 observations in B. We find a 
variability of about 2 magnitudes, with a minimum (B=20.24) at MJD 40760 and a 
second minimum (B=20.19) at MJD 41119. Apart from this minima, the luminosity 
oscillates between B = 18 and B = 19.6 mag (see fig. 7). The object was not seen after 
1985.
%======================= Figure 7 ============================ 
\begin{figure}
\centering
\includegraphics[width=7cm]{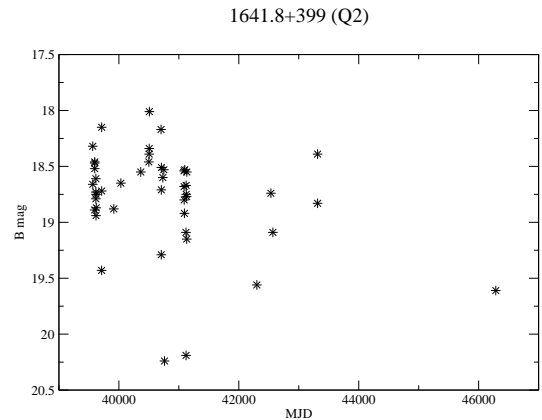} 
\caption{The light curve for 1641.8+399 (Q2)}
\end{figure}
%=============================================================
\subsection{E1641.7+3988 (Q3)}
This object was first identified and spectroscopically confirmed as quasar by 
Crampton et al. (1988, 1989). Never studied for variability, at least to our 
knowledge, this object shows fluctuations of about 1 mag in B between 17.5 and 
18.5 (see Fig. 8). We do not find any indication of violent activity.

%======================= Figure 8 ============================ 
\begin{figure}
\centering
\includegraphics[width=7cm]{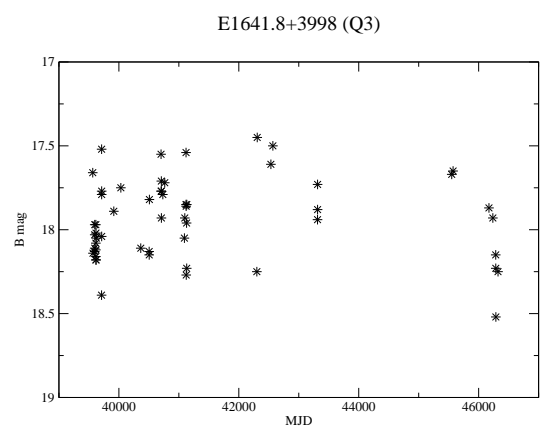} 
\caption{The light curve for E1641.7+3988 (Q3)}
\end{figure}
%=============================================================

\subsection{NGC 6212 (Q4)}

%======================= Figure 9 =========================
\begin{figure}
\centering 
\includegraphics[width=7cm]{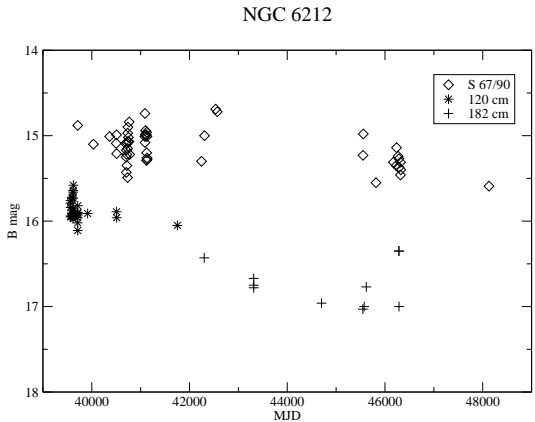} 
\caption{The light curves of NGC 6212. The 3 symbols refer to the 3 used telescopes.}
\end{figure}
%============================================================

Q4 is a type 1 Seyfert galaxy known as NGC 6212, which emits non only in the optical band but 
also in the Radio and in the X bands. A strong outburst (a factor of ten) in X-ray 
was reported by Hailey and Helfand (1980). In 1985 Biermann et al. identified a 
strong variable X-ray source within the galaxy; this variability covers a period 
of one month. 
Burbidge (2003) pointed out that NGC 6212 is surrounded by a large number of QSOs, 
including 3C 345, which lies only 4.7 arcmin away from the centre of the galaxy. 
There are two pairs of QSOs with the same redshifts, including 3C 345, very close to the nucleus of NGC 6212, and the very high surface density of QSOs rapidly falls off with distance from the galaxy. We were aware of the difficulties of performing photometry of a compact nucleus embedded in a fainter extended region on photographic material, and that DAOPHOT is not entirely suitable to that purpose. 
We analyzed separately the light variations from the different telescopes because we expected the mean level of the magnitude to depend in a systematic way on  the numerical aperture of the telescope.
This was actually the case, as can be seen in fig. 9.  The residual variability never exceeds $\pm 0.3$ for the 122 and 182 telescope and $\pm 0.5$ for the S67, and we regard it as highly suspicious.
In order to provide additional confirmation of the reality of the variability we also performed the surface photometry of the galaxy with the Ellipse package of STSDAS. The comparison of the magnitudes of the nuclear region obtained by the two methods shows good agreement because the nuclear region of NGC 6212 is always well visible in our images, even in those obtained by the Schmidt telescope.
At the same time, our results underline the need of further observations of this galaxy in the optical band with linear detectors.

\section{Conclusion} 
We investigated the variability of 3C 345 and of other 4 QSOs in the same
field from the digitized plates of the Asiago archive. While for 3C 345 the variability was already known, for the other three quasars it is reported here for the first time.
The low level variability detected for NGC 6212 is however not beyond doubt.
We underline that these new results have been made possible by the Italian
digitization program, and testify its scientific usefulness.

\section{Acknowledgments}
Thanks are due to an anonymous referee for the careful reading which permitted to improve an earlier version of the paper. Mark Kidger kindly provided the revised photometric sequence of the field.

%========================== Table  =========================
\begin{table*}
\centering 
\begin{minipage}{120mm}
\caption{Log book of the observations}  
\begin{flushleft}
\begin{tabular}{cccccccccc}
\hline
\noalign{\smallskip} 
\noalign{\medskip} 
Plate & Year & MJD & Q1 & Q2 & Q3 & NGC 6212 & 3C 345 & Tel & Filter \\
\noalign{\medskip} 
\hline 
\noalign{\medskip} 
07988 & 1967 & 39564.1458 &  & 18.66 &  & 15.79 & 16.1 & 122 & B\\ 
07989 & 1967 & 39564.1597 & 18.76 & 18.32 & 17.66 & 15.76 & 16.21 & 122 & B\\ 
08004 & 1967 & 39568.0479 &  &  &  &  15.95& 16.33 & 122 & B\\ 
08005 & 1967 & 39568.0646 &  &  &  18.14&  15.84& 16.26 & 122 & B\\ 
08006 & 1967 & 39568.0715 &  &  &  &  15.94& 16.31 & 122 & B\\ 
08030 & 1967 & 39596.984 & 18.83 & 18.46 & 18.11 &15.95  & 15.94 & 122 & B\\ 
08031 & 1967 & 39597.0076 & 18.70 & 18.52 & 18.03 & 15.87 & 15.84 & 122 & B\\ 
08032 & 1967 & 39597.0306 &  & 18.45 & 17.38 & 14.98 & 15.05 & 122 & U\\ 
08033 & 1967 & 39597.0674 & 19.24 & 18.89 & 18.13 & 15.72 & 15.85 & 122 & B\\ 
08035 & 1967 & 39597.1069 & 19.09 & 18.47 & 17.97 & 15.93 & 15.93 & 122 & B\\ 
08036 & 1967 & 39597.1236 &  &  &  & 15.74 & 16.53 & 122 & B\\ 
08039 & 1967 & 39616.0354 & 18.91 & 18.79 & 18.08 & 15.91 & 16.25 & 122 & B\\ 
08040 & 1967 & 39616.0514 & 18.73 & 18.73 & 17.97 & 15.94 & 16.29 & 122 & B\\ 
08041 & 1967 & 39616.0632 & 18.88 & 18.61 & 18.16 & 15.97 & 16.26 & 122 & B\\ 
08042 & 1967 & 39616.0847 & 17.74 & 17.74 & 17.08 & 16.04 & 15.46 & 122 & U\\ 
08043 & 1967 & 39616.1035 &  &  & 18.12 & 15.85 & 16.2 & 122 & B\\ 
8051b & 1967 & 39618.0035 & 18.68 & 18.75 & 18.18 & 15.73 & 16.12 & 122 & B\\ 
08052 & 1967 & 39618.0104 &  &  &  & 15.17 & 15.83 & 122 & V\\ 
08053 & 1967 & 39618.0167 & 18.64 &  &  & 15.66 & 16.06 & 122 & B\\ 
08054 & 1967 & 39618.0243 &  &  &  & 15.05 & 15.96 & 122 & V\\ 
08055 & 1967 & 39618.0333 & 19.16 & 18.94 & 18.18 & 15,64 & 16,03 & 122 & B\\ 
08056 & 1967 & 39618.0486 &  & 17.39 & 17.15 &16.01  & 15.44 & 122 & U\\ 
08057 & 1967 & 39618.0799 &  &  & 18.02 & 15.91 & 16.14 & 122 & B\\ 
08058 & 1967 & 39618.0861 &  &  &  & 15.07 & 15.96 & 122 & V\\ 
08059 & 1967 & 39618.0965 & 17.89 & 17.66 & 17.28 & 16.13 & 15.48 & 122 & U\\ 
08078 & 1967 & 39622.9694 & 19.16 & 18.87 & 18.05 & 15.58 & 16.06 & 122 & B\\ 
08079 & 1967 & 39622.9861 &  &  & 18.42 & 14.38 & 15.88 & 122 & V\\ 
08094 & 1967 & 39710.8646 &  &  & 18.39 & 16.02 & 16.08 & 122 & B\\ 
08096 & 1967 & 39710.8986 &  &  &  & 15.93 & 15.9 & 122 & B\\ 
00810 & 1967 & 39711.934 &  & 19.43 & 17.52 & 14.88 & 16.12 & S67 & B\\ 
08102 & 1967 & 39711.9965 &  &  &  & 16.11 & 16.3 & 122 & B\\ 
08107 & 1967 & 39712.8868 &  &  &  &  & 16.01 & 122 & B\\ 
08108 & 1967 & 39712.9028 &  & 18.15 & 17.79 & 15.91 & 15.91 & 122 & B\\ 
08109 & 1967 & 39712.9125 &  &  & 18.04 & 15.82 & 15.98 & 122 & B\\ 
08110 & 1967 & 39712.9306 & 18.31 & 17.91 & 17.34 & 16.05 & 15.4 & 122 & U\\ 
08111 & 1967 & 39712.9653 & 18.77 & 18.72 & 17.77 & 15.9 & 15.9 & 122 & B\\ 
08112 & 1967 & 39712.9722 &  &  &  & 15.96 & 16.05 & 122 & B\\ 
08295 & 1968 & 39915.1722 &  & 18.88 & 17.89 & 15.91 & 17 & 122 & B\\ 
08296 & 1968 & 39915.1806 &  &  & 18.03 & 14.54 & 17.12 & 122 & V\\ 
01684 & 1968 & 40032.9125 &  & 18.65 & 17.75 & 15.1 & 17.12 & S67 & B\\ 
02397 & 1969 & 40365.0187 &  & 18.55 & 18.11 & 15.01 & 16.43 & S67 & B\\ 
02561 & 1969 & 40499.7646 &  & 18.46 &  & 15.09 & 17.39 & S67 & B\\ 
08597 & 1969 & 40504.7799 &  & 18.39 & 18.13 & 15.89 & 17.23 & 122 & B\\ 
02615 & 1969 & 40505.7736 & 19.04 & 18.34 & 18.15 & 14.99 & 17.17 & S67 & B\\ 
02652 & 1969 & 40508.7646 &  & 18.01 & 17.82 & 15.21 & 17.14 & S67 & B\\ 
08619 & 1969 & 40508.809 &  &  &  & 15.96 & 17.37 & 122 & B\\ 
03279 & 1970 & 40702.9806 &  & 18.17 & 17.55 & 15.25 & 16.73 & S67 & B\\ 
03281 & 1970 & 40705.9389 &  &  & 17.77 & 15.43 & 16.83 & S67 & B\\ 
03290 & 1970 & 40706.9715 &  & 19.29 & 17.77 & 15.1 & 16.65 & S67 & B\\ 
03303 & 1970 & 40707.9062 &  & 18.71 & 17.71 & 15.17 & 16.83 & S67 & B\\ 
03322 & 1970 & 40708.9771 &  & 18.51 & 17.93 & 15.08 & 16.75 & S67 & B\\ 
03330 & 1970 & 40717.0236 &  &  &  & 15.22 & 16.89 & S67 & B\\ 
\noalign{\medskip} 
\hline 
\noalign {\medskip} 
\end{tabular}
\end{flushleft} 
\end{minipage}
\end{table*}

\begin{table*}
\centering 
\begin{minipage}{122mm}
\caption{Log book of the observations}  
\begin{flushleft}
\begin{tabular}{cccccccccc}
\hline
\noalign{\smallskip} 
\noalign{\medskip} 
Plate & Year & MJD & Q1 & Q2 & Q3 & NGC 6212 & 3C 345 & Tel & Filter \\
\noalign{\medskip} 
\hline 
\noalign{\medskip} 
03332 & 1970 & 40718.0375 &  &  &  & 15.35 & 16.96 & S67 & B\\ 
03358 & 1970 & 40731.9271 &  &  &  & 15.49 & 16.59 & S67 & B\\ 
03362 & 1970 & 40732.9146 &  & 18.6 & 17.79 & 15.15 & 16.83 & S67 & B\\ 
03370 & 1970 & 40733.9222 &  &  &  & 14.97 & 16.85 & S67 & B\\ 
03379 & 1970 & 40737.9521 &  &  &  & 15.07 & 14.91 & S67 & B\\ 
03387 & 1970 & 40738.9354 &  &  &  & 14.9 & 16.95 & S67 & B\\ 
03401 & 1970 & 40749.9875 &  & 18.53 &  & 15.02 & 16.93 & S67 & B\\ 
03415 & 1970 & 40760.9236 &  & 20.24 &  & 15.07 & 16.91 & S67 & B\\ 
03424 & 1970 & 40764.9153 &  &  & 17.72 & 14.84 & 16.65 & S67 & B\\ 
03453 & 1970 & 40775.9868 &  &  &  & 15.22 & 16.87 & S67 & B\\ 
04367 & 1971 & 41086.9382 & 18.98 & 18.54 &  & 15.01 & 15.97 & S67 & B\\ 
04375 & 1971 & 41087.9986 & 19.44 & 18.68 &  & 14.74 & 16.02 & S67 & B\\ 
04380 & 1971 & 41088.9778 & 18.88 & 18.8 &  & 15.08 & 16.19 & S67 & B\\ 
04390 & 1971 & 41092.9278 &  & 18.92 & 18.05 & 14.99 & 15.67 & S67 & B\\ 
04399 & 1971 & 41099.0007 & 19.23 & 18.53 & 17.93 & 14.94 & 15.61 & S67 & B\\ 
04423 & 1971 & 41118.9799 & 19.95 & 19.09 & 17.85 & 14.99 & 15.98 & S67 & B\\ 
04424 & 1971 & 41119.9486 &  & 20.19 & 17.54 & 15.29 & 15.86 & S67 & B\\ 
04428 & 1971 & 41121.9694 & 19.73 & 18.67 & 17.86 & 14.96 & 15.94 & S67 & B\\ 
04433 & 1971 & 41122.9236 & 18.82 & 18.75 & 18.27 & 15.2 & 15.97 & S67 & B\\ 
04441 & 1971 & 41129.0181 & 19.63 & 18.77 & 17.96 & 15.01 & 15.79 & S67 & B\\ 
04449 & 1971 & 41131.0174 & 19.21 & 19.15 & 18.23 & 15.26 & 15.7 & S67 & B\\ 
04458 & 1971 & 41132.9917 & 19.43 & 18.55 & 17.85 & 15.28 & 15.46 & S67 & B\\ 
09247 & 1973 & 41752.1819 &  &  &  & 16.05 & 16.76 & 122 & B\\ 
07255 & 1974 & 42244.9833 &  &  &  & 15.3 & 16.58 & S67 & B\\ 
00629 & 1974 & 42300.8333 &  & 19.56 & 18.25 & 16.43 & 16.31 & 182 & B\\ 
07459 & 1974 & 42306.8514 &  &  & 17.45 & 15 & 16.4 & S67 & B\\ 
07934 & 1975 & 42534.9667 & 18.88 & 18.74 & 17.61 & 14.69 & 16.48 & S67 & B\\ 
07970 & 1975 & 42567.9438 & 19.5 & 19.09 & 17.5 & 14.72 & 16.51 & S67 & B\\ 
01577 & 1977 & 43312.9611 &  &  &  & 15.77 & 16.35 & 182 & V\\ 
01578 & 1977 & 43312.9757 &  &  & 17.94 & 16.67 & 16.3 & 182 & B\\ 
01584 & 1977 & 43313.9201 & 19.31 & 18.83 & 17.73 & 16.75 & 16.1 & 182 & B\\ 
01585 & 1977 & 43313.9417 & 19.3 & 18.39 & 17.88 & 16.78 & 16.09 & 182 & B\\ 
02859 & 1981 & 44701.1444 &  &  &  & 16.96 & 15.93 & 182 & B\\ 
03309 & 1983 & 45546.8937 &  &  &  & 17.03 & 15.98 & 182 & B\\ 
12153 & 1983 & 45550.8771 &  &  & 17.67 & 15.23 & 15.77 & S67 & B\\ 
12165 & 1983 & 45555.8368 &  &  &  & 14.98 & 15.79 & S67 & B\\ 
03329 & 1983 & 45573.8438 &  &  & 17.65 & 17 & 15.89 & 182 & B\\ 
03372 & 1983 & 45616.8299 &  &  &  & 16.77 & 15.99 & 182 & B\\ 
12491 & 1984 & 45815.9708 &  &  & 19.74 & 15.55 & 16.35 & S67 & B\\ 
12839 & 1985 & 46172.0417 &  &  & 17.87 & 15.31 & 16.34 & S67 & B\\ 
12850 & 1985 & 46227.9722 &  &  &  & 15.35 & 16.69 & S67 & B\\ 
12856 & 1985 & 46236.9236 &  &  & 17.93 & 15.14 & 16.37 & S67 & B\\ 
12890 & 1985 & 46269.9465 &  &  &  & 15.24 & 16.86 & S67 & B\\ 
03623 & 1985 & 46284.8729 &  &  & 18.15 & 16.35 & 16.52 & 182 & B\\ 
03634 & 1985 & 46285.8778 &  & 19.61 & 18.23 & 16.35 & 16.39 & 182 & B\\ 
03642 & 1985 & 46286.8611 &  &  & 18.52 & 17 & 16.54 & 182 & B\\ 
12903 & 1985 & 46286.8854 &  &  &  & 15.27 & 16.45 & S67 & B\\ 
12911 & 1985 & 46289.8674 &  &  &  & 15.37 & 16.5 & S67 & B\\ 
12975 & 1985 & 46316.8132 &  &  &  & 15.46 & 16.4 & S67 & B\\ 
12982 & 1985 & 46318.8333 &  &  &  & 15.31 & 16.45 & S67 & B\\ 
13000 & 1985 & 46322.8243 &  &  & 18.25 & 15.4 & 16.36 & S67 & B\\ 
14861 & 1990 & 48127.8458 &  &  &  & 15.59 & 16.69 & S67 & B\\ 
\noalign{\medskip} 
\hline 
\noalign {\medskip} 
\end{tabular}
\end{flushleft} 
\end{minipage}
\end{table*}
%===============================================================

\label{lastpage}


\begin{thebibliography}{}
 \bibitem{}
 Angione, R. J., 1971, AJ, 76, 412  
\bibitem{}
 Barbieri, C., Romano, G., di Serego, S., Zambon, M., 1977a, A \&A, 59, 419
\bibitem{}
Barbieri, C., Romano, G., di Serego, S., Zambon, M., 1977b, Nat, 268, 63
\bibitem{}
Barbieri, C. et al., 1988, A\&A., 76, 477
\bibitem{}
 Barbieri, C. et al., 2003, ExA, 15, 29 
\bibitem{}
Biermann, P.; Strom, R.; Bartel, N., 1985, A\&A, 147L, 27B
\bibitem{}
Burbidge, G., 2003, ApJ, 586, 119
\bibitem{}
 Caproni, A., Abraham, Z., 2004, ApJ, 602, 625 
\bibitem{}
 Crampton, D., Cowley, A.P., Schmidtke, P.C., Janson, T., Durrell, P., 1988, AJ, 96, 816
\bibitem{}
 Crampton, D., Cowley, A.P., Hartwick, F.D.A., 1989, ApJ, 345, 59 
\bibitem{}
 de Vaucouleurs, G., 1968, Applied Optics, 7, 1513
\bibitem{}
 de Vries, W.H., Becker, R.H., White, R.L., 2003, AJ, 126, 1217 
\bibitem{}
 Gonzales-Perez, J. N., Kidger, M. R., 2001, AJ, 122, 2055 
\bibitem{}
 Hailey, C. J., Helfand, D. J., 1980, BAAS, 12, 488
\bibitem{}
Hawkins, M. R. S., 2002, MNRAS, 329, 76
\bibitem{}
 Veron-Cetty, M.-P., Veron, P., 2003, A\&A, 412, 399   
\bibitem{}
 Zhang, X., Xie, G.Z., Bai, J.M., 1998, A\&A, 330, 469 
\bibitem{}
 Zhang, X., Xie, G.Z., Bai, J.M., Zhao, G., 2000, A\&SS, 271, 1 

\end{thebibliography}
\end{document}